\documentclass[aps,eqsecnum,nofootinbib,superscriptaddress,showpacs]
{revtex4}
\usepackage[dvips]{graphicx}  
\usepackage{amsmath,amsfonts,bm,amssymb}  
\def\%{\phantom{0}}
\def\be{\begin{eqnarray}}
\def\ee{\end{eqnarray}}
\def\bea{\begin{eqnarray}}
\def\eea{\end{eqnarray}}

\newcommand{\eq}[1]{Eq.~(\ref{eq:#1})}
\newcommand{\sect}[1]{Sec.~\ref{sec:#1}}

\newcommand{\fig}[1]{Fig.~\ref{fig:#1}}
\newcommand{\tabl}[1]{Table~\ref{table:#1}}

\newcommand{\RNAdS}{RN-${\rm AdS}_5\:$}
\def\tilr{\tilde{r}}
\newcommand{\etas}{$\eta/s\:$}
\newcommand{\mft}{\tau_{\rm mft}}
\newcommand{\rela}{\tau_{\rm relax}}
\newcommand{\phipm}{$\Phi_\pm\:$}
\newcommand{\phip}{$\Phi_+\:$}
\newcommand{\phim}{$\Phi_-\:$}

\begin{document}
%
%
%
\title{
Viscosity of gauge theory plasma with a chemical potential
from AdS/CFT correspondence
}
\author{Kengo Maeda}
\email{kmaeda@kobe-kosen.ac.jp}
\affiliation{Department of General Education,
Kobe City College of Technology, Kobe, 651-2194, Japan}
\author{Makoto Natsuume}
\email{makoto.natsuume@kek.jp}
\affiliation{Theory Division, Institute of Particle 
and Nuclear Studies, \\
KEK, High Energy Accelerator Research
Organization, Tsukuba, Ibaraki, 305-0801, Japan}
\author{Takashi Okamura}
\email{okamura@ksc.kwansei.ac.jp}
\affiliation{Department of Physics, Kwansei Gakuin University,
Sanda, 669-1337, Japan}
\date{\today}
\begin{abstract}
We compute the strong coupling limit of the shear viscosity for the ${\cal N}=4$ super-Yang-Mill theory with a chemical potential. We use the five-dimensional Reissner-Nordstr\"{o}m-anti-deSitter black hole, so the chemical potential is the one for the $R$-charges $U(1)_R^3$. We compute the quasinormal frequencies of the gravitational and electromagnetic vector perturbations in the background numerically. This enables one to explicitly locate the diffusion pole for the shear viscosity. The ratio of the shear viscosity $\eta$ to the entropy density $s$ is $\eta/s=1/(4\pi)$ within numerical errors, which is the same result as the one without chemical potential.
\end{abstract}
\maketitle

\section{Introduction}

Gauge/gravity dualities or AdS/CFT dualities (anti-deSitter/conformal field theory) are particularly interesting at finite temperature. One can use the dualities to study the dynamics of gauge theory plasmas or one can use them to address the long-standing issues of black holes.

Finite temperature gauge/gravity dualities were proposed in Refs.~\cite{Witten:1998qj,Witten:1998zw}. At finite temperature, gauge theories are expected to deconfine, so the main evidences of the dualities are rooted in the ones for the deconfinement. It is well-known that there are two order parameters for the deconfinement:
\begin{enumerate}
\item Polyakov loops ($\langle P \rangle \neq 0$ in the plasma phase and $\langle P \rangle = 0$ in the confining phase)
\item Large-$N_c$ behaviors of the free energy ($F \sim O(N_c^2)$ in the plasma phase and $F \sim O(1)$ in the confining phase)
\end{enumerate}
As is evident from the nature of these evidences, there are some shortcomings for the finite temperature dualities:
\begin{enumerate}
\item[(i)] The dualities are understood only qualitatively. (This is due to the lack of supersymmetry.)
\item[(ii)] Thermodynamic properties are compared,%
\footnote{Klebanov, with a series of collaborators, has pioneered such  comparison even before Refs.~\cite{Witten:1998qj,Witten:1998zw}.
See, {\it e.g.}, Refs.~\cite{Gubser:1996de,Klebanov:1996un,Klebanov:1997kc,Gubser:1998bc}.}
but it is not evident that the dualities hold even for nonequilibrium processes.
\item[(iii)] Gauge theories other than ${\cal N}=4$ super-Yang-Mills theory (SYM) are less understood. For zero temperature, many backgrounds with less supersymmetry are known. But even those $T=0$ backgrounds are rather involved, so their $T \neq 0$ generalization is a formidable task (See Refs.~\cite{Buchel:2003tz,Buchel:2004hw,Benincasa:2005iv,Buchel:2005cv} and references therein for some exceptions.)
\end{enumerate}

Remarkably, there is a claim to solve these problems all at once, which is the hydrodynamic description of gauge theory plasmas using gauge/gravity dualities \cite{Buchel:2003tz}-\cite{Kovtun:2005ev}. It is originally studied in the context of ${\cal N}=4$ SYM \cite{Policastro:2001yc} and later is generalized to many cases. The claim is that gauge theory plasmas which have gravity duals have a universal  ratio of the shear viscosity $\eta$ to the entropy density $s$ at large 't~Hooft coupling:
\be
\frac{\eta}{s} =\frac{\hbar}{4\pi k_B}.
\ee
This result overcomes the above difficulties in the following sense: (i) The result is a definite quantitative statement; (ii) Nonequilibrium processes are considered; (iii) It is expected to be satisfied for a wide range of gauge theories. Interestingly, RHIC (Relativistic Heavy Ion Collider) has observed the ``elliptic flow" \cite{rhic}, which may be a consequence of the shear viscosity, and the estimated value is rather close to the duality values \cite{Teaney:2003pb,Hirano:2005wx}.

The hydrodynamic description is powerful yet there are many open issues. Of particular interest in view of experiments is the case with a chemical potential. All discussion so far is limited to zero chemical potential. The general theorems of the universality fail in the presence of a chemical potential \cite{Kovtun:2003wp,Buchel:2003tz,Kovtun:2004de,Buchel:2004qq}. However, real experiments are done in the finite baryon number density, so one has to consider systems with a chemical potential. 

It is not clear how one can incorporate the baryon number density in the present scheme. However, a charged black hole is a simple way to introduce a chemical potential since the electromagnetic potential plays the role of a chemical potential for black holes. In this paper, we use the five-dimensional Reissner-Nordstr\"{o}m-AdS (RN-${\rm AdS}_5$) black hole to study the issue.

Our main tool is the numerical computation of the quasinormal (QN) frequencies of the black hole. This approach has been used to compute the shear viscosity, {\it e.g.}, in Refs.~\cite{Nunez:2003eq,Kovtun:2005ev}.
In hydrodynamics, the shear viscosity accounts for the diffusion. In black hole physics, the diffusion is due to the QN modes.

This is not an easy task however since the perturbation equations are highly-coupled. When the number of degrees of freedom are increased, the task is extremely hard. (Thus, many authors often consider the gravitational tensor perturbation %
\footnote{See \sect{QNM} for the definition of various perturbation modes.}; 
the tensor perturbation reduces to a minimally-coupled scalar field. Then, one uses the Kubo formula to derive the viscosity.) Fortunately, the perturbation equations in question are studied by Kodama and Ishibashi \cite{Kodama:2003kk}; they have decoupled the differential equations, so we use their formalism to obtain the shear viscosity. We compute the gravitational vector perturbation and locate the pole explicitly for the shear viscosity. For completeness, we also compute the electromagnetic vector perturbation, which is technically very similar to the gravitational vector perturbation.

The plan of the present paper is as follows. First, in the next section, we summarize some background materials; We give the relevant physics of the \RNAdS black hole and discuss the relationship between the QN frequencies and hydrodynamics. In \sect{method}, we present the perturbation equations and a numerical method to obtain the QN frequencies. In \sect{results}, our results are discussed. We conclude in \sect{discussion} with discussion and implications.

After this work is being completed, we learned that Refs.~\cite{Mas:2006dy,Son:2006em,Saremi:2006ep} also compute the shear viscosity for charged AdS black holes. We briefly discuss their results and compare with our results at the end of this paper. 

\section{Backgrounds}

\subsection{Five-dimensional RN-AdS black hole}\label{sec:RN-AdS}

The \RNAdS black hole is a solution of the five-dimensional Einstein-Maxwell-AdS theory:
\begin{equation}
  S = \frac{1}{16 \pi G} \int d^5x \sqrt{-g}
      \left[ R +\frac{12} {l^2} \right]
    - \frac{1}{4} \int d^5x \sqrt{-g} F^2
\label{eq:action}
\end{equation}
(We use the unit $8 \pi G = 1$ and set the AdS radius $l=1$.) The black hole (with noncompact horizon) is given by
\bea
  ds^2 &=& - f(r) dt^2 + f^{-1}(r) dr^2 + r^2 (dx_1^2+dx_2^2+dx_3^2)~,
\\
  f(r) &=& r^2 - 2M/r^2 + Q^2/r^4
  = \frac{ ( r^2 - r_+^2 ) ( r^2 - r_-^2 ) ( r^2 + r_0^2 ) }{r^4}~,
\label{eq:RNAdS}
\eea
where $r_0^2 = r_+^2 + r_-^2$.
The black hole with noncompact horizon corresponds to a large black hole limit of the \RNAdS black hole with $S^3$ horizon.

The surface gravity $\kappa$ and the mass parameter $M$ are
written by $r_-/r_+$ as
\begin{align}
  & \frac{\kappa}{r_+}
  = 2 - \left( \frac{r_-}{r_+} \right)^2
  - \left( \frac{r_-}{r_+} \right)^4~,
\\
  & \frac{2\, M}{r_+^4}
  = 1 + \left( \frac{r_-}{r_+} \right)^2
  + \left( \frac{r_-}{r_+} \right)^4~.
\end{align}

In order to see the gauge theory interpretation of the charge, it is convenient to see the embedding of the black hole into ten dimensions. 
The original finite temperature duality is
\begin{center}
${\cal N}=4$ SYM at finite temperature $\leftrightarrow$ type IIB string theory in Schwarzschild-${\rm AdS}_5$ black holes (${\rm SAdS}_5$) $\times S^5$
\end{center}
One can add an angular momentum along $S^5$, which is known as the ``spinning" D3-brane solutions \cite{Kraus:1998hv,Cvetic:1999xp}. The angular momentum becomes a charge after the $S^5$ reduction. The symmetry of $S^5$ corresponds to an internal symmetry of SYM, R-symmetry $SO(6)$. The R-symmetry group $SO(6)$ is rank 3, so one can add at most three independent charges. The three-charge solution is known as the STU solution \cite{Behrndt:1998jd}. When all charges are equal, the STU solution is nothing but the \RNAdS black hole. In this case, the charge corresponds to the diagonal $U(1)$ of the $U(1)_R^3$ charges.

Because the charge corresponds to the $U(1)_R$ charge, this is by no means realistic. However, the theory has interesting features which are common to the real QCD. For instance, the phase diagram \cite{Chamblin:1999tk,Cvetic:1999ne} is qualitatively similar to the QCD diagram.%
\footnote{This is the case of the RN-AdS black hole with compact horizon or compact SYM. This paper considers the RN-AdS black hole with noncompact horizon, which is always in plasma phase.}
Our system is not a realistic chemical potential, but it may mimic the realistic case and one may learn an interesting lesson for the case of chemical potential.

It is straightforward to compute various physical quantities of the black hole using the Euclidean path integral technique \cite{Chamblin:1999tk}. According to gauge/gravity dualities, they are translated into the corresponding quantities of a gauge theory (the energy density $\epsilon$, the pressure $p$, the temperature $T_{\text{H}}$, the entropy density $s$, the chemical potential $\Phi_{\text{H}}$, and the charge density $\mathcal{Q}$):
\begin{align}
  & \epsilon = 3\, M~,
& & p = M~,
& & T_{\text{H}} = \frac{\kappa}{2 \pi}~,
& & s = 2\, \pi\, r_+^3~,
  & \Phi_{\text{H}} = \sqrt{\frac{3}{2}}~\frac{Q}{r_+^2}~,
& & \mathcal{Q} = \sqrt{6}\, Q~.
\label{eq:thermo}
\end{align}
The speed of sound is then given by 
$v_s=\sqrt{\partial p/\partial \epsilon}=1/\sqrt{3}$,
which results from a scale invariance of the solution~(\ref{eq:RNAdS}). (The symmetry is part of conformal invariance at zero temperature.) The invariance also requires the bulk viscosity to vanish $\zeta=0$.

\subsection{QN frequencies and hydrodynamics}\label{sec:QNM}

The QN frequencies measure how black hole perturbations decay. (See Ref.~\cite{Kokkotas:1999bd} for a comprehensive review.) Since the initial value acts as a source, the main interest is the retarded Green's functions with appropriate boundary conditions. It turns out that the Green's functions have a number of complex poles in the complex $\omega$-plane, which means that the perturbations decay exponentially in time. So, the diffusion is mainly governed by the lowest pole.

To be more specific, let us consider the gravitational perturbations which take the form $h_{\mu\nu}(r) \, e^{-i\omega t + ikx_3}$. We look for the lowest pole which survives in the low-frequency, long-wavelength limit $\omega(k)\rightarrow 0$, $k\rightarrow 0$ (hydrodynamic limit). The diffusion is due to the black hole absorption. One interprets the diffusion as a consequence of viscosities of the dual gauge theory plasmas.

The perturbations can be decomposed by the world-volume symmetry $O(3)$ of $x_1, \cdots, x_3$. (We follow the conventions of Ref.~\cite{Kodama:2003kk}.%
\footnote{This convention is slightly different from the one of Ref.~\cite{Kovtun:2005ev}. In Ref.~\cite{Kovtun:2005ev}, the perturbations $h_{\mu\nu}(r) e^{-i\omega t +iqx_3}$ are considered and they are decomposed by the remaining world-volume symmetry $O(2)$ of $x_1$ and $x_2$.})
Then, the gravitational perturbations are decomposed as the tensor mode, the vector mode (``shear mode"), and the scalar mode (``sound mode").%
\footnote{In Ref.~\cite{Kovtun:2005ev}, these modes are called the spin~2, spin~1, spin~0 modes, respectively.}

Such a decomposition is essentially the same as hydrodynamics. In hydrodynamics, the vector mode measures the shear viscosity, and the scalar mode measures the bulk viscosity and the speed of sound. For example, as we will see below, the vector mode satisfies the following dispersion law (in the hydrodynamic limit):
\begin{align}
  & \frac{\omega}{\kappa}
  = - i~D_\eta~\left( \frac{k}{\kappa} \right)^2~,
& & D_\eta := \frac{ \kappa~\eta }{\epsilon + p}~.
\label{eq:dispersion}
\end{align}
Then, one can regard $\eta$ as the shear viscosity of the dual gauge theory plasma.

Similarly, the electromagnetic perturbations are decomposed as the vector mode and the scalar mode. In this paper, we compute the QN frequencies for the vector modes since we are mainly interested in the shear viscosity. 

As is evident from the above discussion, this is a Lorentzian computation. On the other hand, gauge/gravity dualities are originally defined in the Euclidean signature (such as the GKP-Witten relation \cite{Witten:1998qj,Gubser:1998bc}). Many authors have studied the Lorentzian prescription \cite{Balasubramanian:1998sn,Balasubramanian:1998de,Balasubramanian:1999ri,Son:2002sd} and we essentially use the one given in Ref.~\cite{Son:2002sd}. In principle, one can use only the Euclidean prescription and analytically continue the results to the Lorentzian signature. However, the differential equations in gravity backgrounds often have more than three regular singularities, so the analytic computations are not possible; one can locate only the pole locations numerically. Without knowing the full analytic properties, the analytic continuation is not possible. So, one has to work with the Lorentzian signature. In simple cases where the analytic solutions are available (such as the three-dimensional black hole), the Euclidean prescription and the Lorentzian prescription by Ref.~\cite{Son:2002sd} give the same results.

\section{Numerical method}\label{sec:method}


The perturbative equations for the Einstein-Maxwell-AdS system (\ref{eq:action}) have been studied in Ref.~\cite{Kodama:2003kk}. (See Sec.~4.3.1 of the paper for the vector perturbations, which are our primary interests.) Let us outline their procedure briefly. 
One first has to find the gauge-invariant variables from the gravitational and electromagnetic perturbations. 
Denote the gauge-invariant gravitational and electromagnetic vector perturbations expanded in vector harmonics by $\Omega$ and ${\cal A}$, respectively. 
The equations of $\Omega$ and ${\cal A}$ themselves are not very useful however since they are highly coupled. But take the following linear combinations of the gauge-invariant variables:
\footnote{The system studied by Kodama and Ishibashi is the general Einstein-Maxwell system with a cosmological constant in $d=n+2$ dimensions. Set $n=3$,  $K=0$ (the spatial curvature of the horizon), and $\lambda=-1$ (the sign of the cosmological constant) in their equations. Also, see the paper for the definition of the numerical coefficients $a_\pm$ and $b_\pm$.}
\be
\Phi_\pm = a_\pm r^{-3/2}\Omega + b_\pm r^{1/2} {\cal A}.
\label{eq:variables}
\ee
Then, the differential equations decouple from each other. In the limit of zero charge, the fields $\Phi_-$ and $\Phi_+$ reduce to the pure gravitational mode and pure electromagnetic mode, respectively. 

Let us use the dimensionless coordinates $\tilr=r/r_+$ and $\tilr_-=r_-/r_+$. The decoupled equations are then given by 
\begin{align}
\label{original-eq}
  & 0 = \left( f(\tilr)\, \frac{d}{d\tilr}\, f(\tilr)\, \frac{d}{d\tilr}
  + \omega^2 - V(\tilr) \right)~\Phi_\pm(\tilr)~,
\\
  & V(r) = \frac{f(\tilr)}{\tilr^2}\, \left[~k^2 + \frac{3}{4}\, \tilr^2
  + \frac{39}{4}\, \frac{Q^2}{\tilr^4} + \frac{\mu_\pm}{\tilr^2}~\right]~,
\\
  & \mu_\pm := - \frac{11}{2}\, M \pm \Delta~,
& & \Delta := \sqrt{ 64 M^2 + 12 k^2 Q^2 }~,
\end{align}
where $2 M = 1 + \tilr_-^2 + \tilr_-^4$ and~$Q^2 = \tilr_-^2 ( 1 + \tilr_-^2 )$.

The QN modes are the solutions of the perturbative equations with appropriate boundary conditions. The boundary condition at the outer-horizon is the purely ingoing mode, which is physically natural for the retarded Green's functions. The boundary condition at infinity is chosen to be a decaying mode. The asymptotic solutions behave as $\Phi \sim \tilr^{-3/2}$ or $\Phi \sim \tilr^{1/2}$, and the former one is chosen by regularity.
Incorporating the boundary conditions both for the horizon and infinity, 
we obtain the form 
\be
\label{form}
\Phi_\pm(\tilr)=\left( \frac{\tilr^2 - 1}{\tilr^2 - \tilr_-^2} \right)^{- i \omega/2 \kappa}
    \tilr^{-3/2}\psi_\pm(\tilr).
\ee
The equation~(\ref{original-eq}) is transformed into the form 
\begin{align}
  & 0 = \left( s(z)\, z^2\, \frac{d^2}{dz^2}
  + t(z)\, z\, \frac{d}{dz} + u(z) \right)~\psi_\pm(z)~,
\end{align}
where $z=(\tilr^2 - 1)/(\tilr^2 - \tilr_-^2)$ and
\begin{align}
   s(z)
  &= 4 ( -1 + \tilr_-^2 ) ( 2 + \tilr_-^2 )^2 ( -1 + z )
  ( -1 + \tilr_-^2 z ) \big\{ -2 + z + \tilr_-^2 ( -1 + 2 z )~\big\}^2~,
\\
\nonumber \\
   t(z)
  &= 4 ( 2 + \tilr_-^2 ) \big\{ -2 + z + \tilr_-^2 ( -1 + 2 z )~\big\}
\nonumber \\
  & \times \Bigg[~( -1 + \tilr_-^2 ) ( 2 + \tilr_-^2 )
  \bigg( -2 - 4 ( -2 + z ) z + 2 \tilr_-^4 z ( -1 + z + z^2 )
\nonumber \\
  & \hspace{3.5cm}
  + \tilr_-^2 \Big[ -1 + z \big\{~3 + ( -7 + z ) z~\big\}~\Big]~\bigg)
\nonumber \\
  & \hspace{0.5cm}
  + i\, \omega\, ( -1 + z ) ( -1 + \tilr_-^2 z )
     \big\{ -2 + z + \tilr_-^2 ( -1 + 2 z )~\big\}~\Bigg]~,
\\
\nonumber \\
   u(z)
  &= z \Bigg[ - ( -1 + \tilr_-^2 ) ( 2 + \tilr_-^2 )^2
  \big\{ -2 + z + \tilr_-^2 ( -1 + 2 z )~\big\}
\nonumber \\
  &\hspace*{3.0cm} \times \Big(~k^2 ( -1 + \tilr_-^2 z )
  + ( -1 + z ) \big\{~4 ( 1 + \tilr_-^2 + \tilr_-^4 ) \pm \Delta~\big\}~\Big)
\nonumber \\
  & \hspace*{0.0cm}
  + 2 i\, \omega\, ( -2 + \tilr_-^2 + \tilr_-^4 )
      \big\{ -2 + z + \tilr_-^2 ( -1 + 2 z )~\big\}
      \big\{ -5 + 3 z + \tilr_-^2 ( -3 + 5 z )~\big\}
\nonumber \\
  & \hspace*{0.0cm}
  + \omega^2\, ( -1 + \tilr_-^2 z )
    \Big(~8 + ( -5 + z ) z + \tilr_-^8 z^2 + \tilr_-^6 z ( -3 + 5 z )
         + 3 \tilr_-^4 \big\{~1 + z ( -5 + 3 z )~\big\}
\nonumber \\
  & \hspace*{3.0cm}
         + \tilr_-^2 \big\{~10 + z ( -19 + 5 z )~\big\}~\Big)~\Bigg].
\end{align}
The solutions are expanded as a series to solve the differential equation:
\be
\psi_\pm(\tilr)=\sum_{n=0}^\infty a_nz^n.
\ee
Then, the expansion coefficients $a_n$ satisfy the following 
recurrence relations: 
\be 
\label{recurrence}
&&\alpha_0 a_1+\beta_0 a_0=0, \nonumber \\
&&\alpha_1 a_2+\beta_1 a_1+\gamma_1 a_0=0, \nonumber \\
&&\alpha_2 a_3+\beta_2 a_2+\gamma_2 a_1+\delta_2 a_0=0, \nonumber \\
&&\alpha_n a_{n+1}+\beta_n a_n
+\gamma_n a_{n-1}+\delta_n a_{n-2}+\rho_n a_{n-3}=0, \qquad 
n=3,\,4,\, \cdots,
\label{eq:recurrence}
\ee
where the recurrence coefficients, which depend on $\omega$,
are given in appendix.
The purely ingoing wave boundary condition at the horizon
is reflected by the first equation in (\ref{eq:recurrence}),
and the fall-off condition $\Phi \sim \tilde{r}^{-3/2}$ at infinity
is satisfied for the series $\psi_\pm(z)$ which converges uniformly as $z \rightarrow 1$
(namely, for the series where $\sum a_n$ exists and is finite.)
Both conditions are satisfied by special complex values of $\omega$
which are the QN frequencies.

The convergence problem of the series translates into the convergence of the continued fraction (in a compact notation of continued fractions)
\begin{align}
   \frac{a_1}{a_{0}}
  &= \frac{- \gamma'_1}{ \beta'_1 - {~} }\,
       \frac{ \alpha'_1 \gamma'_{2} }
            { \beta'_{2} - {~} }\,
       \frac{ \alpha'_{2} \gamma'_{3} }
            { \beta'_{3} - \cdots }~.
\end{align}
Here, $\{\alpha'_n\}, \{\beta'_n\}$, and $\{\gamma'_n\}$~~
($n = 0, 1, 2, \cdots$)
are the coefficients of the three term recurrence relation
\be 
&&\alpha'_0 a_1+\beta'_0 a_0=0, \nonumber \\
&&\alpha'_n a_{n+1}+\beta'_n a_n+\gamma'_n a_{n-1} =0, \qquad 
n=1,\,2,\cdots,
\ee
which is obtained from the original recurrence relation
(\ref{eq:recurrence}) by the Gaussian elimination.
Thus, the QN frequencies are obtained by solving the equation 
\cite{Gautschi:1967}
\begin{align}
   - \frac{\beta'_0}{\alpha'_0}
  &= \frac{- \gamma'_1}{ \beta'_1 - {~} }\,
       \frac{ \alpha'_1 \gamma'_{2} }
            { \beta'_{2} - {~} }\,
       \frac{ \alpha'_{2} \gamma'_{3} }
            { \beta'_{3} - \cdots }~.
\end{align}

\section{Results}\label{sec:results}

We compute the QN frequencies for $\Phi_\pm$, which are linear combinations of the gravitational and electromagnetic vector perturbations. 
We first discuss general structure of these poles and thier gauge theory interpretations. Then, we closely study the pole which survives in the hydrodynamic limit. 

\subsection{General QN spectrum}

%
\begin{figure*}[htb]
  \begin{center}
       \includegraphics[width=7.5cm,height=5.0cm,clip]{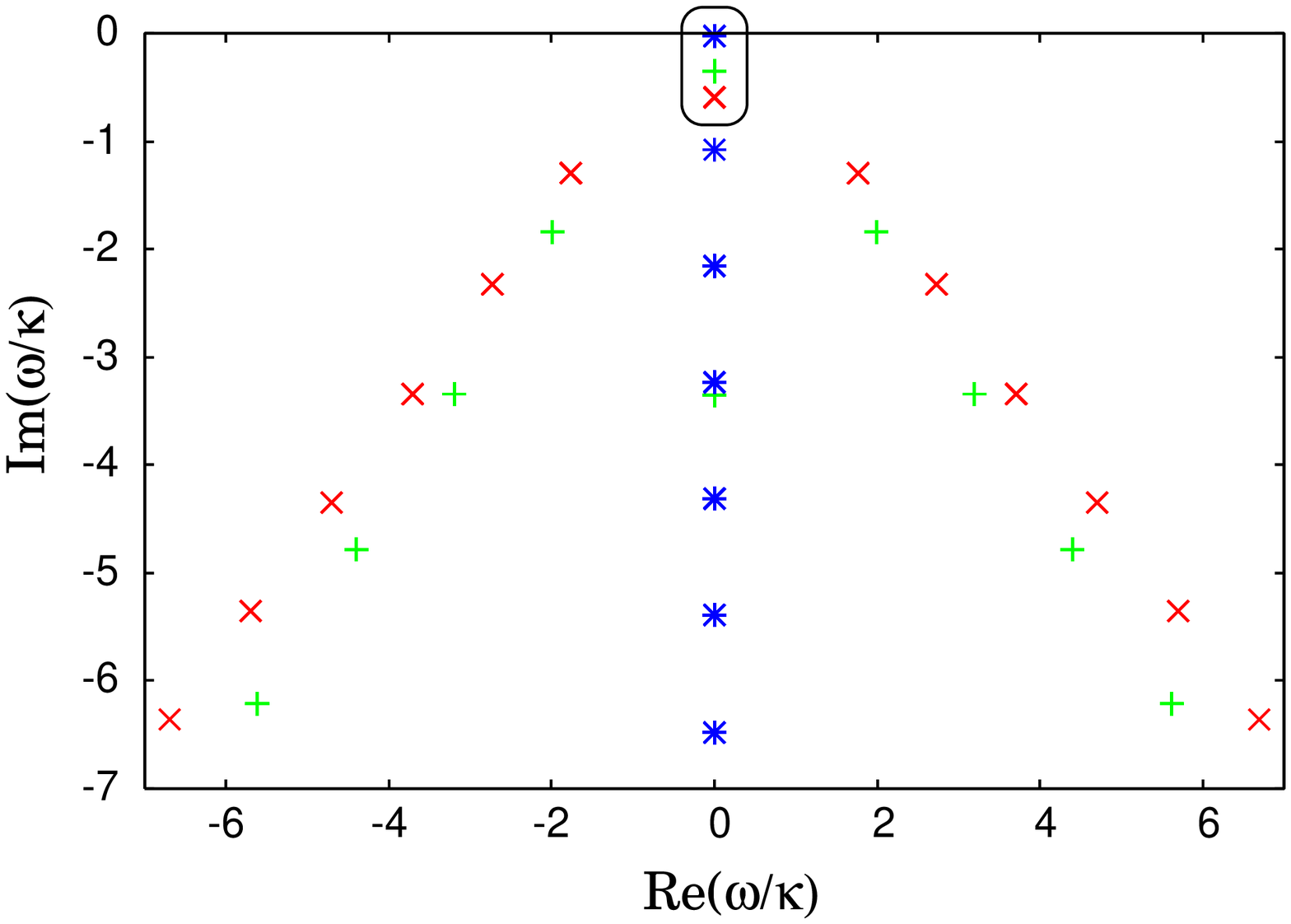}
     \hfil
       \includegraphics[width=7.5cm,height=5.0cm,clip]{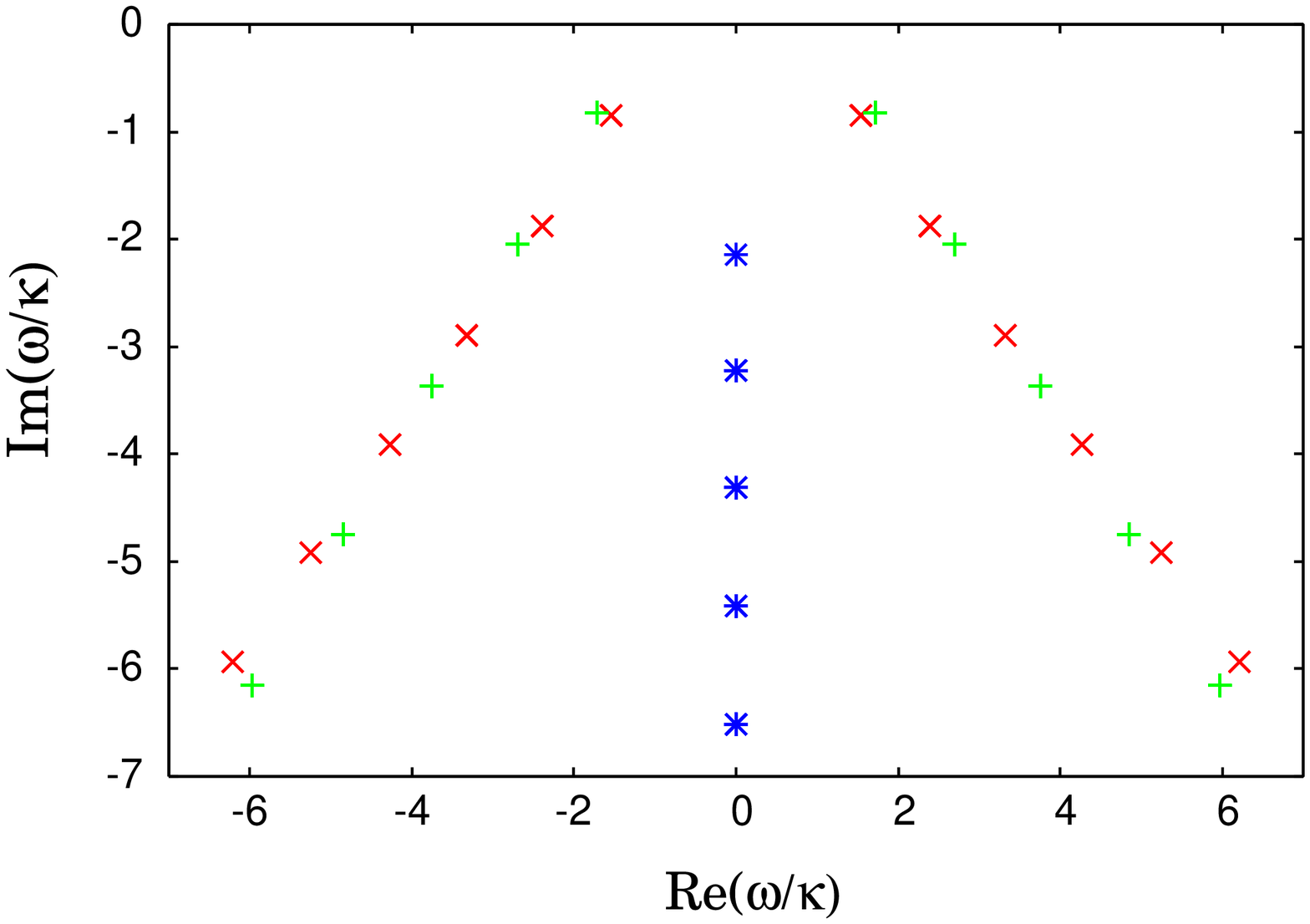}
  \end{center}
\caption{\label{fig:V-_rm-dep_k_100}
(color online). 
QN spectrum of the perturbations \phim (left panel) and \phip (right). Dependence of $\omega/2 \pi T_{\text{H}}$
on $r_-/r_+ = 0$($\times$),~$0.5$($+$), and $0.95$ ($\divideontimes$) is shown 
for spatial momentum $k/2 \pi T_{\text{H}} = 1.0$.
On the left panel, the poles which are enclosed by the elongated circle are the hydrodynamic pole discussed in \sect{hydro}. In contrast, there is no such poles for $\Phi_+$. All poles stay at a finite distance from the real axis.}
\end{figure*}%
%

The poles for the perturbations \phipm are shown in \fig{V-_rm-dep_k_100} (\phim on the left panel and \phip on the right panel). 
On the left panel, the poles which are enclosed by the elongated circle are the hydrodynamic pole; they approach the real axis in the limit $\omega(k)\rightarrow 0$, $k\rightarrow 0$. We closely study the pole in \sect{hydro}. In contrast, \phip does not have a pole which survives in the hydrodynamic limit. All poles stay at a finite distance from the real axis even in the limit. 

Let us look at gauge theory interpretation briefly.
According to the dictionary of gauge/gravity dualities, the gravitational field couples to the energy-momentum tensor of the gauge theory, so the gravitational perturbations correspond to the energy-momentum tensor correlators. 
The electromagnetic field couples to the $R$-current, so the electromagnetic perturbations correspond to $R$-current correlators in the dual gauge theory side. 

In our choice of field variables $\Phi_\pm$, the gravitational and  electromagnetic perturbations are entangled with each other [See \eq{variables}]. So, our computation mixes the energy-momentum tensor correlator with the $R$-current correlator. In general, one has to go back to the original variables to extract information of gauge theory correlators. However, in this case, such a disentanglement is not really necessary in order to know the plasma viscosity. This is because \phip has no hydrodynamic pole. The dispersion relation of the shear mode is simply determined by the hydrodynamic pole of $\Phi_-$. This also means that the vector mode of the $R$-current correlator has a hydrodynamic pole with the same dispersion relation as the energy-momentum tensor correlator. [Namely, there is a pole which satisfies the dispersion relation (\ref{eq:dispersion}) with the same diffusion constant as $D_\eta$.]

The poles beyond the hydrodynamic pole are not relevant to hydrodynamics but are interesting from the point of view of gauge/gravity dualities. One feature of these poles is that they lie approximately linearly. The slope is getting sharper as $r_-$ is increased. 
This property that poles lie linearly has been observed for the SAdS black holes (see, {\it e.g.}, Refs.~\cite{Nunez:2003eq,Kovtun:2005ev,Starinets:2002br,Cardoso:2003cj})
and for the RN-${\rm AdS}_4$ black hole \cite{Berti:2003ud}.
This is also observed for the near-horizon limit of nonextreme D$p$-branes \cite{Maeda:2005cr}. 
However, its origin and significance are far from obvious.

\subsection{Hydrodynamic pole}\label{sec:hydro}

%
\begin{table}[htb]
\begin{center}
\begin{tabular*}{12cm}{@{\extracolsep{\fill}}ccccc}
\hline\hline
   $~~r_-/r_+$ & $a$ & $b$ & $c \times 10^{11}$
  & $\gamma_{\text{num.}}$
\\ \hline
   $~~0.0\%$ & $-0.500051\%$ & $2.00002$ & $-3.08339\%\%$
 & $1.000102\%\%\%$~
\\
   $~~0.05$ & $-0.498175\%$ & $2.00002$ & $-2.7581\%\%\%$
 & $1.00010035\%$~
\\
   $~~0.1\%$ & $-0.49255\%\%$ & $2.00002$ & $-2.94714\%\%$
 & $1.00010002\%$~
\\
   $~~0.15$ & $-0.483179\%$ & $2.00002$ & $-2.51994\%\%$
 & $1.00009449\%$~
\\
   $~~0.2\%$ & $-0.470089\%$ & $2.00002$ & $-2.54955\%\%$
 & $1.0000913\%\%$~
\\
   $~~0.25$ & $-0.453336\%$ & $2.00002$ & $-2.33166\%\%$
 & $1.00008669\%$~
\\
   $~~0.3\%$ & $-0.433032\%$ & $2.00001$ & $-1.86945\%\%$
 & $1.00007874\%$~
\\
   $~~0.35$ & $-0.409367\%$ & $2.00001$ & $-1.62585\%\%$
 & $1.0000732\%\%$~
\\
   $~~0.4\%$ & $-0.382616\%$ & $2.00001$ & $-1.39078\%\%$
 & $1.0000651\%\%$~
\\
   $~~0.45$ & $-0.353155\%$ & $2.00001$ & $-1.17109\%\%$
 & $1.00006151\%$~
\\
   $~~0.5\%$ & $-0.321446\%$ & $2.00001$ & $-0.971175\%$
 & $1.00005422\%$~
\\
   $~~0.55$ & $-0.288032\%$ & $2.00001$ & $-0.791317\%$
 & $1.00004974\%$~
\\
   $~~0.6\%$ & $-0.253502\%$ & $2.00001$ & $-0.628083\%$
 & $1.00004391\%$~
\\
   $~~0.65$ & $-0.218464\%$ & $2.00001$ & $-0.478805\%$
 & $1.00003943\%$~
\\
   $~~0.7\%$ & $-0.183507\%$ & $2.00001$ & $-0.343142\%$
 & $1.00003295\%$~
\\
   $~~0.75$ & $-0.149172\%$ & $2.00001$ & $-0.223444\%$
 & $1.00002414\%$~
\\
   $~~0.8\%$ & $-0.115927\%$ & $2.0\%\%\%\%$ & $-0.126007\%$
 & $1.00001675\%$~
\\
   $~~0.85$ & $-0.0841502$ & $2.0\%\%\%\%$ & $-0.0553647$
 & $1.00001171\%$~
\\
   $~~0.9\%$ & $-0.0541242$ & $2.0\%\%\%\%$ & $-0.0130611$
 & $1.00000517\%$~
\\
   $~~0.95$ & $-0.0260391$ & $2.0\%\%\%\%$ & $\%0.0549852$
 & $0.999999434$~
\\ \hline\hline
\end{tabular*}
\end{center}
\caption{Fitting parameters and $\gamma_{\text{num.}}$
for various charge-to-mass ratios.
With a good accuracy, $b=2$, $c=0$, and $\gamma_{\text{num.}}=1$.}
\label{table:fitting_param}
\end{table}
%

%
\begin{figure*}[htb]
  \begin{center}
       \includegraphics[width=12.0cm,clip]{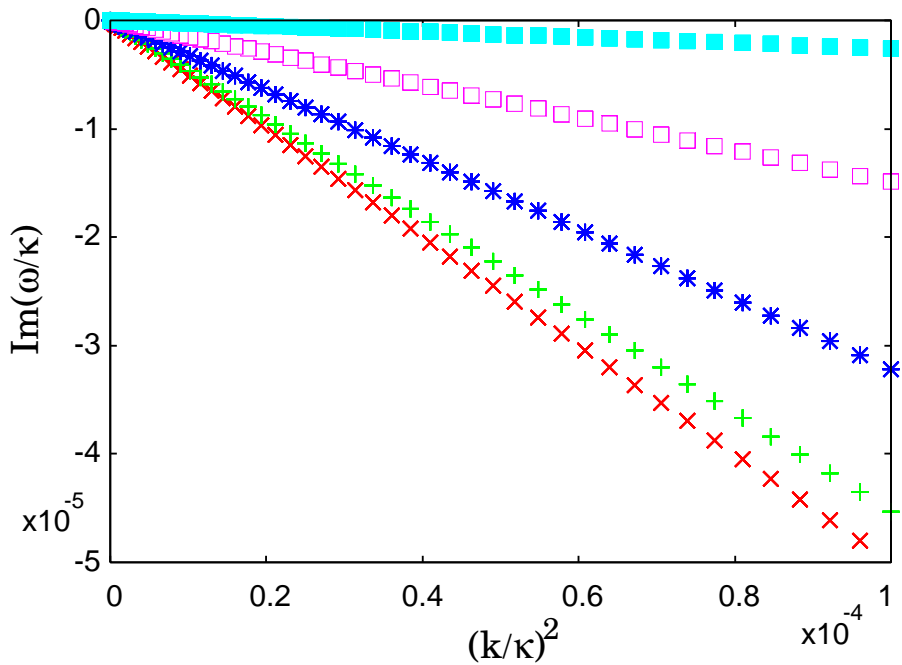}
  \end{center}
\caption{\label{fig:k-dep}
(color online). 
Dependence of $\omega_I/2 \pi T_{\text{H}}$
on $(k/2 \pi T_{\text{H}})^2$
for $r_-/r_+ = 0$($\times$),~$0.25$($+$),~
$0.5$({\small $\divideontimes$}),~$0.75$({\small $\boxdot$}),~and 
$0.95$({\small $\blacksquare$}).
The slope determines the diffusion constant $D_\eta$.}
\end{figure*}%
%
%
\begin{figure*}[htb]
  \begin{center}
       \includegraphics[width=7.5cm,height=5.0cm,clip]{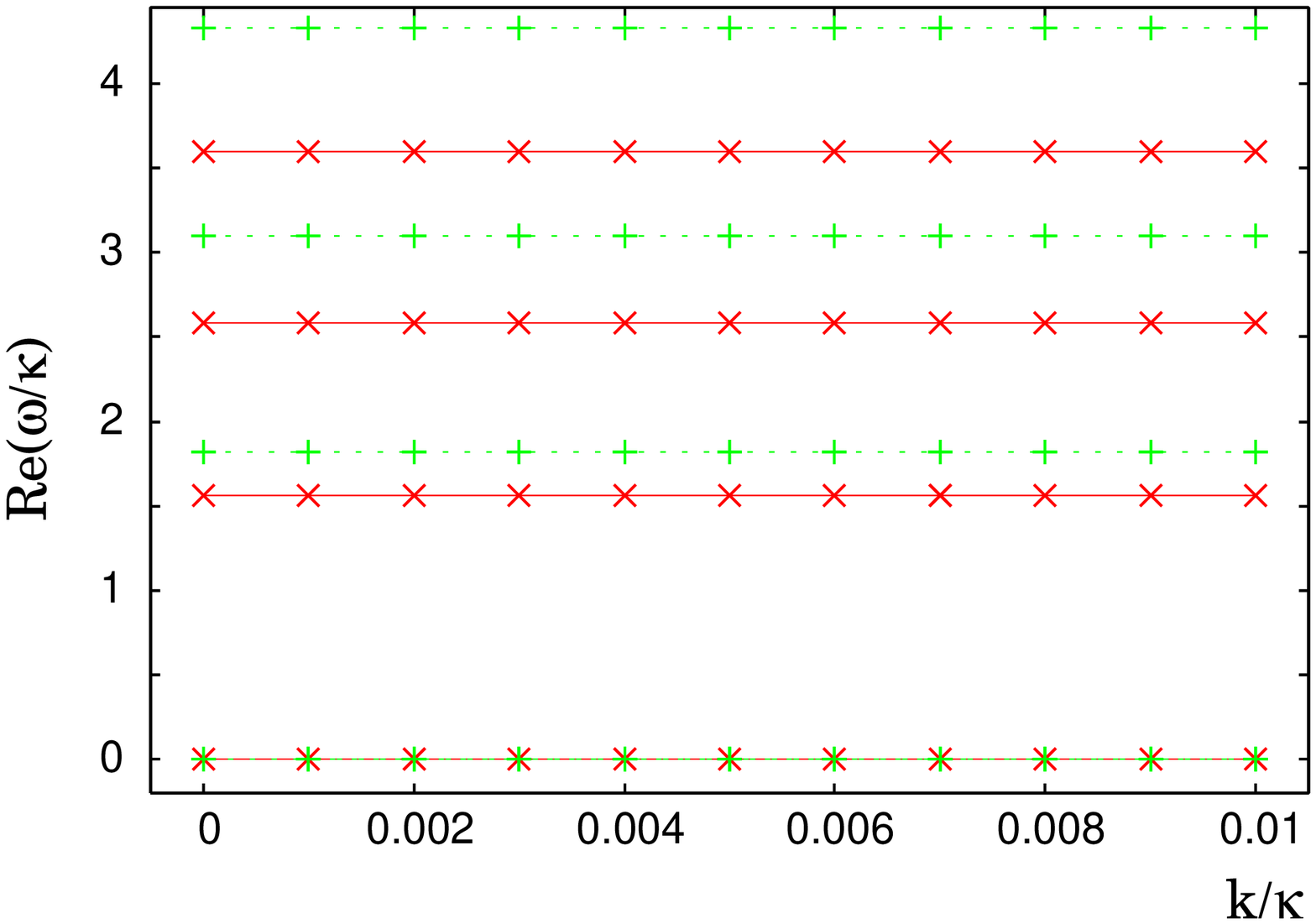}
     \hfil
       \includegraphics[width=7.5cm,height=5.0cm,clip]{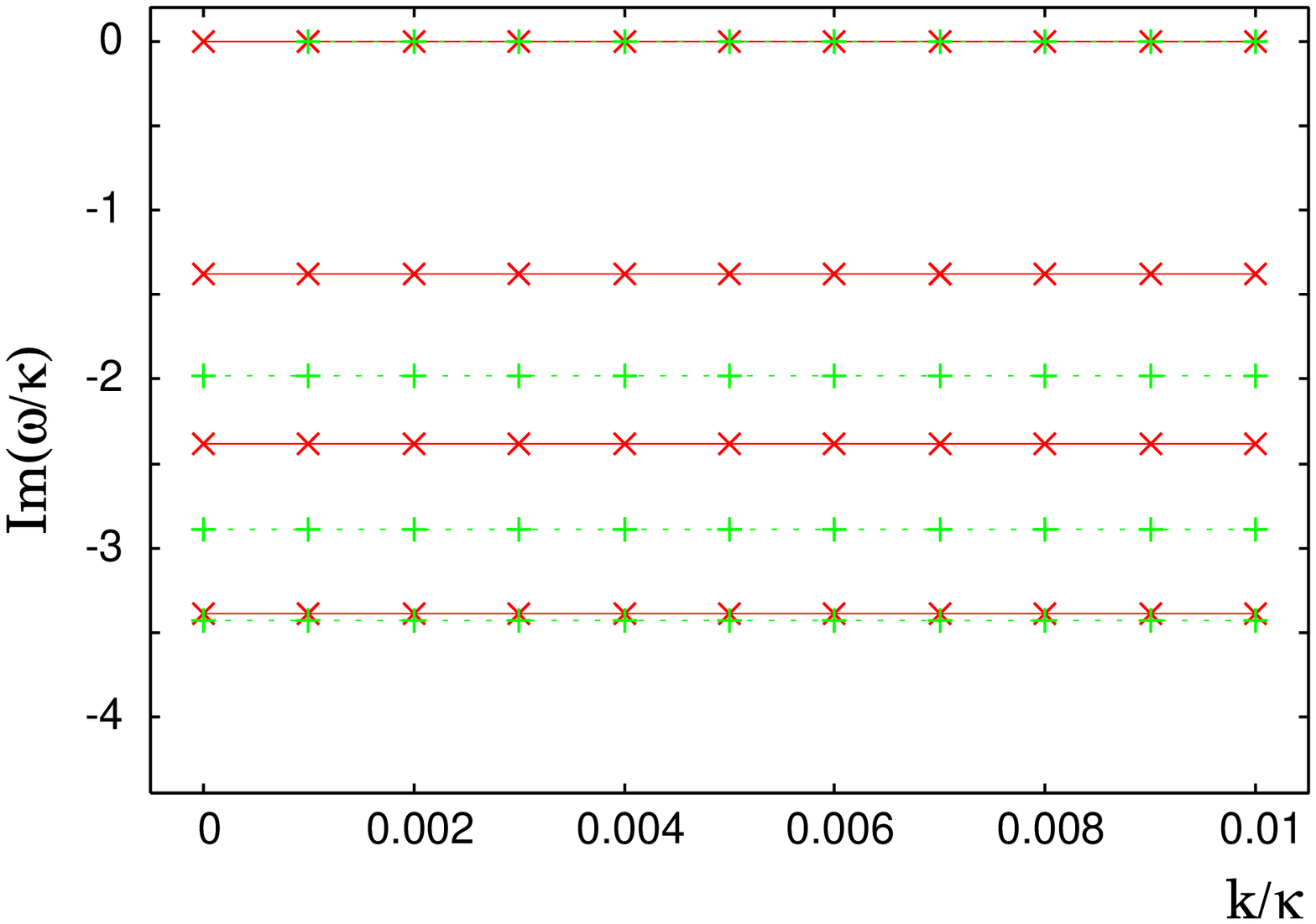}~
  \end{center}
\caption{\label{fig:slope_k-dep}
(color online). 
Dependence of four low-lying QNM's 
on $k/2 \pi T_{\text{H}}$ for $r_-/r_+ = 0$($\times$),~$0.5$($+$).
The left panel shows the real part of $\omega/2 \pi T_{\text{H}}$
and the right shows the imaginary part. 
The hydrodynamic pole is the one which approaches $\omega\rightarrow 0$ as $k \rightarrow 0$ (the bottom one on the left panel and the top one on the right panel). As one can see, the other poles stay at a finite distance from the origin.}
\end{figure*}%
%

Let us look at the hydrodynamic mode of \phim in more detail. 
In the hydrodynamic limit $\omega(k)\rightarrow 0$, $k\rightarrow 0$, the gravitational vector mode (shear mode) is expected to satisfy the following dispersion law:
\begin{align}
  & \frac{\omega}{\kappa}
  = - i~D_\eta~\left( \frac{k}{\kappa} \right)^2~,
& & D_\eta := \frac{ \kappa~\eta }{\epsilon + p}~.
\label{eq:shear_diffusion}
\end{align}
We numerically estimate the diffusion constant $D_\eta$ through the QN frequencies. The ratio $\eta/s$ can be written as
\begin{align}
   \frac{\eta}{s}
  &= \frac{\epsilon + p}{ \kappa~s }~D_\eta
  =: \frac{1}{4 \pi}~\gamma~,
\label{eq:viscosity-bound} \\
   \gamma
  &:= \frac{2 (\epsilon + p)}{ T_{\text{H}}~s }~D_\eta
  = \frac{ 4~\big[~1 + (r_-/r_+)^2 + (r_-/r_+)^4~\big] }
         { 2 - (r_-/r_+)^2 - (r_-/r_+)^4 }~D_\eta~,
\label{eq:def-gamma}
\end{align}
where \eq{thermo} is used. The quantity $\gamma$ is \etas in units of $1/(4\pi)$. We estimate $\gamma$ below; $\gamma=1$ means that \etas is the same as the one without the chemical potential.

The diffusion constant $D_\eta$ is estimated by choosing the following fitting function:
\begin{align}
  & \frac{\omega_I}{\kappa}
  = a \times \left( \frac{k}{\kappa} \right)^b + c~,
\label{eq:fitting_formula}
\end{align}
where $a,~b$ and $c$ are the fitting parameters.
Then, the parameter $a$ is the numerically estimated value
of $D_\eta$ and the numerically calculated $\gamma$ is given by
\begin{align}
   \gamma_{\text{num.}}
  &:= \frac{ 4~\big[~1 + (r_-/r_+)^2 + (r_-/r_+)^4~\big] }
         { 2 - (r_-/r_+)^2 - (r_-/r_+)^4 }~a~.
\label{eq:def-gamma_num}
\end{align}

The result is summarized in \tabl{fitting_param} and Figs.~\ref{fig:k-dep}, \ref{fig:slope_k-dep}. As one can see from the last column of the table, $\gamma_{\text{num.}}=1$ with a good accuracy, which means that
\be
\frac{\eta}{s} = \frac{1}{4\pi}
\ee
even in the presence of a chemical potential. This is our main result.

\section{Discussion}\label{sec:discussion}

We found that \etas is again $1/(4\pi)$ in the presence of a chemical potential. A nonzero value of viscosity is probably not very surprising for fluids at finite temperature. There is an intuitive argument why the viscosity bound exists \cite{Kovtun:2004de}. Generically, the viscosity is given by $\eta \sim \epsilon \mft$, where $\epsilon$ is the energy density and $\mft$ is the mean free time. The entropy density is roughly the order of the number density $n$. So, \etas is the order of average energy of the particle times the mean free time. From the uncertainty relation, this product must be larger than $\hbar$. 

This argument is an intuitive argument, but this suggests that nonzero viscosity is also expected in the presence of a chemical potential (although one cannot learn the precise value from the argument). 
Our computation confirms the naive expectation and \etas is the same as the one without a chemical potential.

Such a dimensional argument should also hold for the other transport coefficients. So, one would expect some relations among various coefficients. (In some cases, one would expect a bound similar to the viscosity bound.) In fact, the existence of the relations among various coefficients is well-known in the context of the linear response theory.

In thermodynamic description, the viscosity bound also constrains the behavior of the relaxation time $\rela$. From $\eta \sim \epsilon \mft$ and $\epsilon \sim n T$, one obtains $\eta/s \sim T \mft$. So, the bound implies that $\mft$ is independent of the coupling in the strong coupling limit and depends only on the temperature. Since $\rela>\mft$, the saturation of $\mft$ mean that $\rela$ cannot be small indefinitely either. 

In fact, Ref.~\cite{Horowitz:1999jd} points out that the relaxation time is saturated at strong coupling and is the order of thermal wavelength. The relaxation time becomes shorter as the coupling is increased, but cannot be smaller than the time scale determined by the thermal wavelength and causality. In gravity description, this is a consequence of a scale invariance (which is part of conformal invariance at zero temperature) for the large black hole limit of SAdS black holes. The noncompact $(p+1)$-dimensional SYM, which is described by the near-horizon limit of D$p$-branes in gravity side, are no longer conformal, but this is also true for these systems \cite{Maeda:2005cr}.

The hydrodynamic description using gauge/gravity duals has many open issues. For example, there are several independent methods to calculate the viscosity such as the Kubo formula  arguments, ``membrane paradigm"-like approach, and the direct computation of the QN frequencies. Currently, these three gave the same results (for the cases where the comparison is possible), which justify these approaches, but the relationship is not clear. Basically, these methods can be classified into two groups:
\begin{enumerate}
\item Phenomenologically, the viscosity is given by the one-point correlator of the energy-momentum tensor.
\item Microscopically, the viscosity is given by the two-point correlators of the energy-momentum tensor, namely by the Kubo formula.
\end{enumerate}
These two methods are of course not independent. Rather, they are related by a linear response theory. Thus, one may be able to establish the connection among these methods if one really establishes the linear response theory for AdS black holes. This in turn may give the nonequilibrium verification of finite-temperature gauge/gravity dualities.

After this work is being completed, we learned that three other groups also compute the shear viscosity for charged AdS black holes  \cite{Mas:2006dy,Son:2006em,Saremi:2006ep}. For the convenience of the readers, we summarize their results and the differences from our paper. All four groups have obtained $\eta/s=1/(4\pi)$ even with chemical potentials, but the systems and the techniques are slightly different from each other:
\begin{itemize}
\item Ref.~\cite{Mas:2006dy} studied the five-dimensional STU black hole. The gravitational tensor perturbation is computed analytically and the shear viscosity is determined by the Kubo formula. 
\item Ref.~\cite{Son:2006em} studied the five-dimensional STU black hole. The gravitational tensor perturbation is computed analytically and the shear viscosity is determined by the Kubo formula. For the STU solution with a single charge, the lowest QN mode of the gravitational vector perturbation is also computed analytically.
\item Ref.~\cite{Saremi:2006ep} studied the RN-${\rm AdS}_4$ black hole. The lowest QN mode of the gravitational vector perturbation is computed numerically (for a small chemical potential). The shear viscosity is expanded as a power series in the chemical potential, and no correction was found for the first few series.
\item We studied the RN-${\rm AdS}_5$ black hole. The QN modes of the gravitational and electromagnetic vector perturbations are computed numerically. 
\end{itemize}

\begin{acknowledgments}
We would like to thank K. Itakura, O. Morimatsu, and T. Sakai for discussions. M.N.\ would like to thank CQUeST (Center for Quantum Spacetime) at Sogang University and Hanyang University for the kind hospitality where part of the paper was written. The research of M.N.\ was supported in part by the Grant-in-Aid for Scientific Research (13135224) from the Ministry of Education, Culture, Sports, Science and Technology, Japan.
\end{acknowledgments}

\appendix*
\section{Recurrence coefficients}

The recurrence coefficients in \eq{recurrence} are given by
\begin{align}
\alpha_n
  &= -4 ( 1 + n ) ( 2 + \tilr_-^2 )^3 ( \kappa + n \kappa - i\, \omega )~,
\\
\nonumber \\
   \beta_n
  &= ( 2 + \tilr_-^2 ) \Bigg[~\kappa\, ( 2 + \tilr_-^2 )
\nonumber \\
  & \hspace*{2.0cm} \times
  \Big[~12 + k^2 + 4 n
    \big\{~5 + 4 n + ( -2 + 7 n ) \tilr_-^2 + ( -3 + n ) \tilr_-^4~\big\}
    \pm \Delta - 4 \kappa~\Big]
\nonumber \\
  & \hspace*{1.5cm}
  - 2 i\, \omega\, \big\{~2 n ( 2 + \tilr_-^2 ) ( 4 + 7 \tilr_-^2 + \tilr_-^4 )
  + \kappa\, ( 5 + 3 \tilr_-^2 )~\big\} - \omega^2\, ( 4 + 3 \tilr_-^2 )~\Bigg]~,
\\
\nonumber \\
   \gamma_n
  &= - \kappa\, ( 2 + \tilr_-^2 ) \Bigg(~k^2 ( 1 + 4 \tilr_-^2 + \tilr_-^4 )
  + 4 \Big[~3 + n + 5 n^2 + \big\{~16 + n ( -29 + 22 n )~\big\} \tilr_-^2
\nonumber \\
  & \hspace*{3.0cm}
  + ( -1 + n ) ( -37 + 22 n ) \tilr_-^4
  + ( -1 + n ) ( -16 + 5 n ) \tilr_-^6~\Big]
\nonumber \\
  & \hspace*{2.0cm}
  + 3 ( 1 + \tilr_-^2 ) \pm \Delta - 12 ( 1 + \tilr_-^2 ) \kappa~\Bigg)
\nonumber \\
  &+ 2 i\, \omega\, \Big[~\kappa\, ( 11 + 26 \tilr_-^2 + 11 \tilr_-^4 )
  + 2 ( -1 + n ) ( 1 + \tilr_-^2 ) ( 2 + \tilr_-^2 )
  ( 5 + 17 \tilr_-^2 + 5 \tilr_-^4 )~\Big]
\nonumber \\
  &+ \omega^2\, ( 5 + 27 \tilr_-^2 + 25 \tilr_-^4 + 6 \tilr_-^6 )~,
\\
\nonumber \\
   \delta_n
  &= \kappa\, ( 2 + 5 \tilr_-^2 + 2 \tilr_-^4 )
  \Big[~4 + ( 96 + k^2 ) \tilr_-^2 \pm \Delta  - 4 \kappa
\nonumber \\
  & \hspace*{2.5cm}
  + 4 \big\{~( -1 + n ) n + n ( -26 + 7 n ) \tilr_-^2
  + ( -2 + n ) ( -13 + 4 n ) \tilr_-^4~\big\}~\Big]
\nonumber \\
  &- 2 i\, \omega\, ( 1 + 2 \tilr_-^2 )
  \big\{~2 ( -2 + n ) ( 2 + \tilr_-^2 ) ( 1 + 7 \tilr_-^2 + 4 \tilr_-^4 )
  + ( 3 + 5 \tilr_-^2 ) \kappa~\big\}
\nonumber \\
  &- \omega^2\, ( 1 + 4 \tilr_-^2 + 2 \tilr_-^4 ) ( 1 + 6 \tilr_-^2 + 2 \tilr_-^4 )~,
\\
\nonumber \\
   \rho_n
  &= \tilr_-^2~\Big[
   - 4 \kappa\, ( -3 + n )^2 ( 2 + \tilr_-^2 ) ( 1 + 2 \tilr_-^2 )^2
\nonumber \\
  & \hspace*{1.5cm}
  + 4 i\, \omega\, ( -3 + n ) ( 2 + \tilr_-^2 ) ( 1 + 2 \tilr_-^2 )^2
  + \omega^2\, ( 1 + 5 \tilr_-^2 + 9 \tilr_-^4 + 5 \tilr_-^6 + \tilr_-^8 )~\Big]~.
\end{align}


\end{document}